\documentclass{sf2a-conf}
\usepackage{graphicx}
\usepackage{natbib}
\def\pasp{PASP}
\def\aap{A\&A}

\def\mnras{MNRAS}

\begin{document}

\TitreGlobal{SF2A 2008}

\title{B[e] stars at the highest angular resolution:\\
  the case of HD87643}
\author{Millour, F.}\address{Max-Planck Institut f\"ur
  Radioastronomie, Auf dem H\"ugel 69, 53121, Bonn, Germany}
\author{Chesneau, O.}\address{UMR 6525 H. Fizeau, Univ. Nice Sophia
  Antipolis, CNRS, Observatoire de la C\^{o}te d'Azur, F-06108 Nice
  cedex 2, France }
\author{Borges Fernandes, M.$^2$}
\author{Meilland, A.$^1$}
%
\runningtitle{HD87643 at the highest angular resolution}
\setcounter{page}{237}

\index{Millour, F.}
\index{Chesneau, B.}
\index{Borges Fernandes, M.}
\index{Meilland, A.}

\maketitle
\begin{abstract}
  New results on the B[e] star HD87643 are presented here. They were
  obtained with a wide range of different instruments, from wide-field
  imaging with the WFI camera, high resolution spectroscopy with the
  FEROS instrument, high angular resolution imaging with the adaptive
  optics camera NACO, to the highest angular resolution available with
  AMBER on the VLTI. We report the detection of a companion to HD87643
  with AMBER, subsequently confirmed in the NACO data. Implications of
  that discovery to some of the previously difficult-to-understand
  data-sets are then presented.
 \end{abstract}
%
\section{Introduction}

B[e] stars (or ``stars with the B[e] phenomenon'') have
little similarities with the classical Be stars. They do show
permitted emission lines from Hydrogen and metallic elements, and they
have an infrared excess. However, they also exhibit forbidden emission
lines, and their infrared excess is due to hot dust ($T\approx
1300-1500$K) instead of heated plasma. Another difference is that the
evolutionary status of B[e] stars is unclear. Some of them show
characteristics typical of evolved objects (e.g. supergiant B[e] stars
or SgB[e]), while others show characteristics of young stellar
objects (Herbig B[e] stars are an example). Hence, ``stars with the
B[e] phenomenon'' do not form a homogeneous class of objects. They
have sometimes a poorly known distance, making their evolutionary
status highly controversial. As a consequence, many B[e] stars are
``classified'' as ``unclassified B[e] stars'' (or UnclB[e]). One
example is the star HD87643, located in the direction of the Carina
arm, which has been classified as SgB[e] due to its putative
distance $\geq1$\,kpc, and which, at the same time, shows variability
typical of a young stellar object. It shows one of the most extreme
infrared excess of all B[e] stars and, hence, needs to be further
investigated. We have observed this star with a variety of techniques
to try to fix its properties and evolutionary status. These techniques
range from high-resolution spectroscopy to high angular resolution
imaging. We present here the high angular resolution discovery images
of a companion to HD87643, as well as a larger scale image of arcs in
its surrounding nebula, that might be related to the binary
star. These results were presented in details in \citet{Millour09}. We
will also present a tentative new interpretation of previously
published spectro-astrometric measurements \citep[from][that failed to
detect the binary]{2006MNRAS.367..737B}, having in mind, now, that
HD87643 is indeed a binary system.

\section{A companion star detected with the highest angular
  resolution.}

\subsection{AMBER + NACO: detection of a companion star}

\begin{figure}[htbp]
  \centering
  \begin{tabular}{cc}
    \multicolumn{2}{c}{\includegraphics[width=0.9\textwidth]{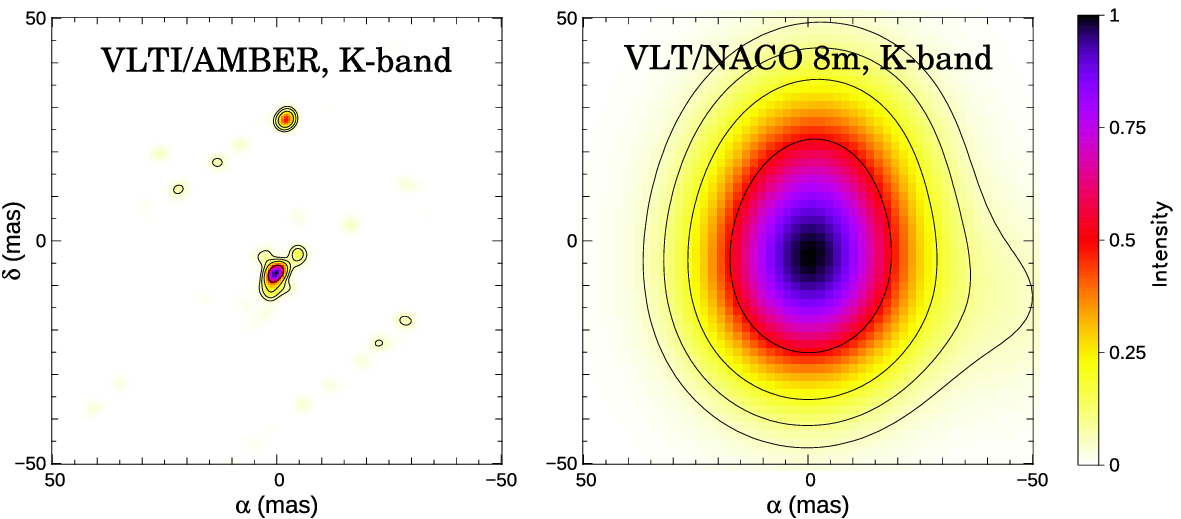}}\\
    \includegraphics[width=0.4\textwidth]{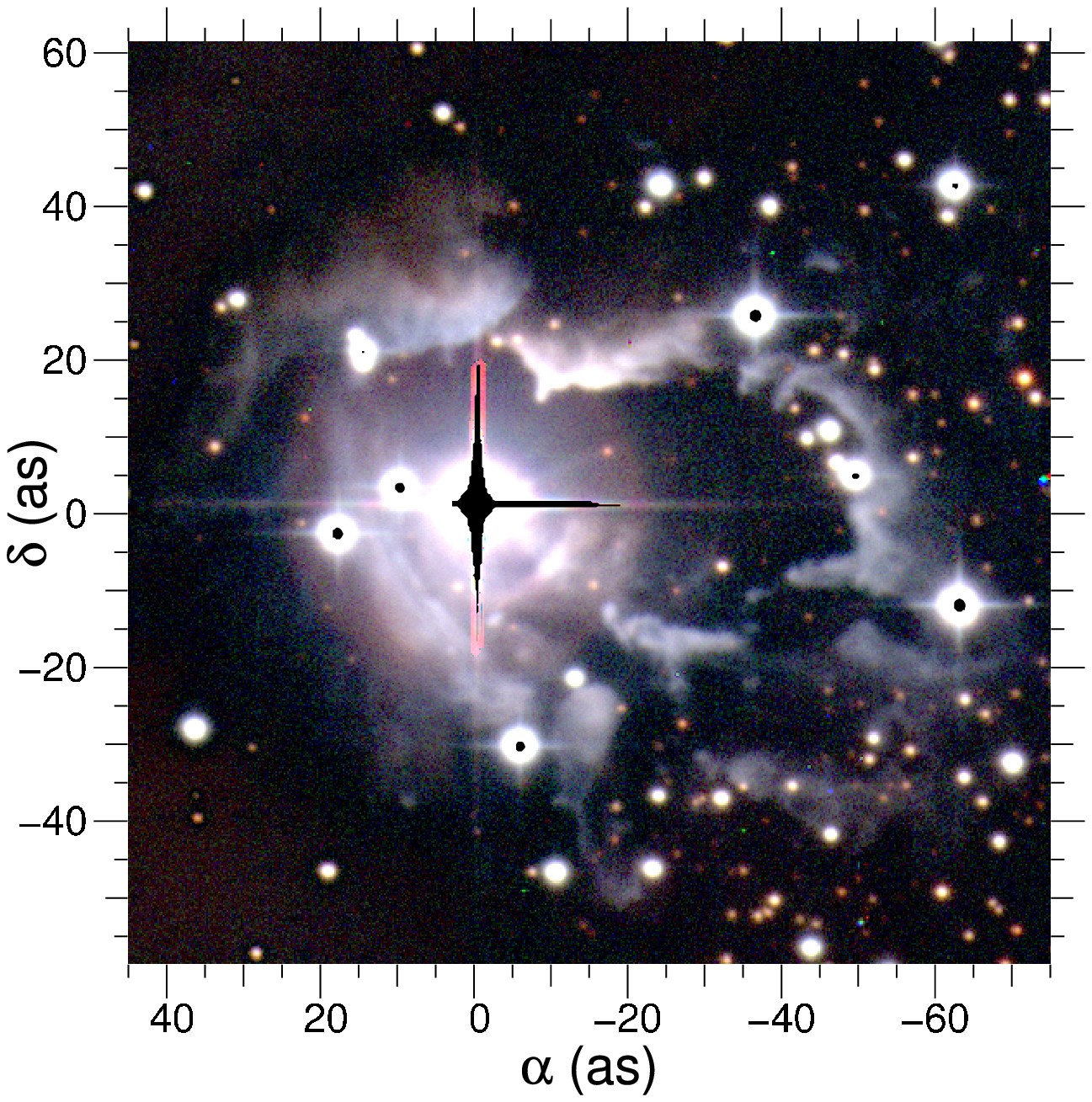}&
    \includegraphics[width=0.4\textwidth, bb=70 0 618 548]{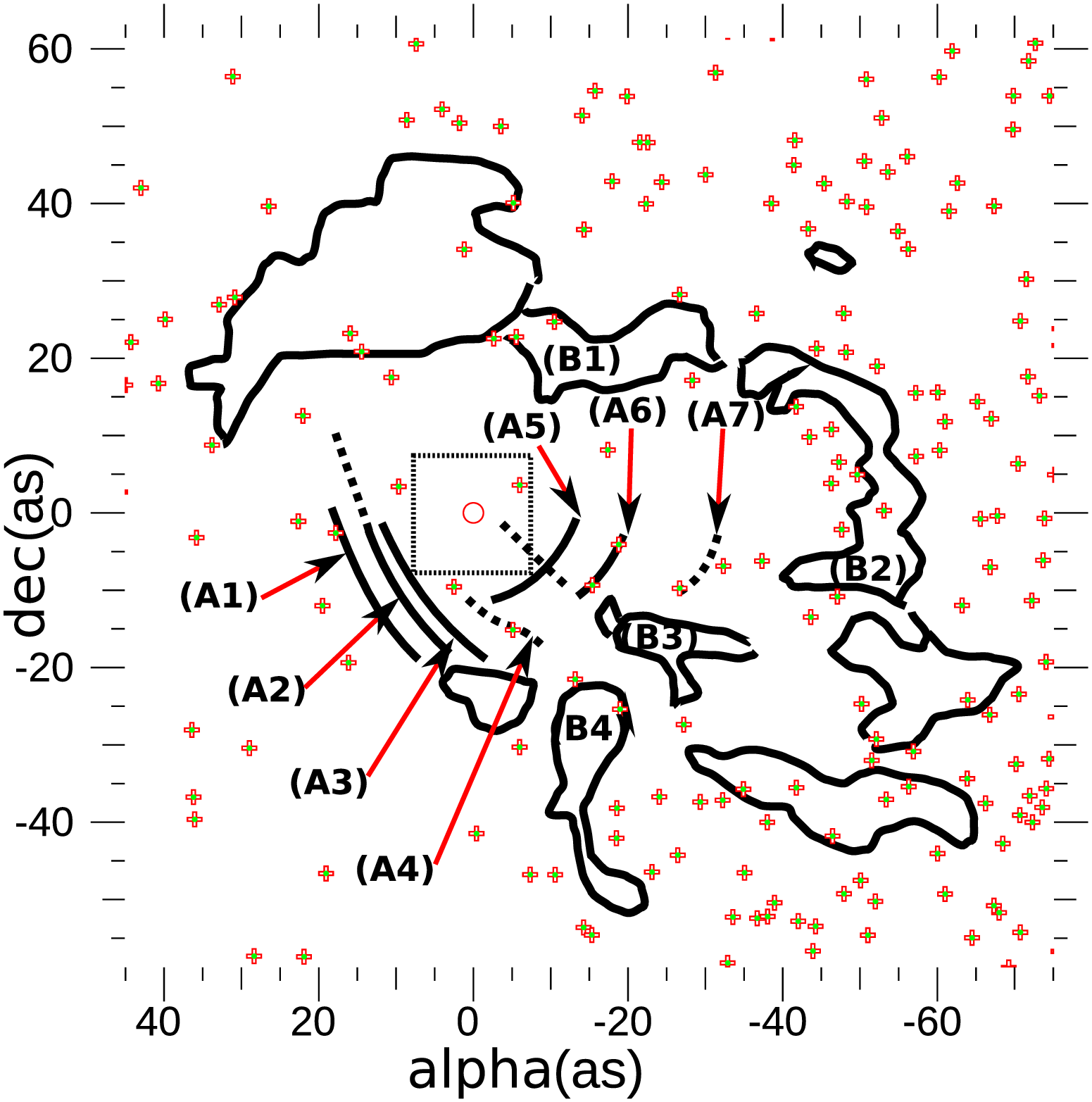}
  \end{tabular}
  \caption{{\bf Top:} High angular resolution images of HD87643: Left:
    AMBER image giving a typical resolution of 2\,mas; Right:
    De-convolved NACO image with a resolution of $\approx50$\,mas.
    {\bf Bottom:} The reflection nebula around HD\,87643: Left: WFI
    image (composite of R, V and B filter); Right: Sketch presenting
    the main structures. Saturated regions are masked by black
    zones. The sketch shows HD\,87643 as a red circle and other stars
    as red crosses. The nebular contours are drawn as black lines, and
    the prominent features are labelled (A1) to (A7) for arc-like
    structures and (B1) to (B4) for knots. Faint or uncertain
    structures are marked as dotted lines.
  }
  \label{figure_Hires_images}
\end{figure}

We observed HD87643 using AMBER \citep[][]{2007A&A...464....1P}, the
near-infrared instrument of
the VLTI in 2006 and in 2008. The AMBER data-sets were recorded in the
K-band in medium spectral resolution ($R=1500$) in 2006 and at low
spectral resolution ($R=35$) in 2008. We were able to perform an image
reconstruction analysis with various image reconstruction software:
MIRA, BSMEM, and BBM. All gave the same result, i.e. the clear
separation of the source into two components and the partial
resolution of the southern component (see
Fig.~\ref{figure_Hires_images}, top-left). A subsequent test using a
home-made model-fitting tool gave basically the same result. This
probably means that a companion star has been detected around HD87643
with AMBER.

To check this discovery, we got additional observing time with the
adaptive optics camera NACO \citep{2003SPIE.4839..140R} of the VLT. We
found that the L-band images did not show any elongation, as the
angular resolution of NACO at that wavelength does not allow the
partial resolution of the binary. On the contrary, the K-band
de-convolved image was barely elongated in the same direction as the
binary star detected with AMBER (see Fig.~\ref{figure_Hires_images},
top-right). Therefore, the NACO observation fully supports the AMBER
observation, which detected a companion star for HD87643.

\subsection{MIDI+AMBER: characterizing the system}

In addition, we observed HD87643 in 2006 in low spectral resolution
($R=30$) using MIDI \citep[][]{2004A&A...423..537L}, the mid-infrared
instrument of the VLTI. Since MIDI only provides visibilities and
differential phases, it is virtually impossible to recover an image
from its data. Thanks to the same model-fitting software as described
before, we could indeed fit the MIDI data-sets with a binary star
model, whereas previous attempts using a 2D radiative-transfer dusty
disk model completely failed to do so.

We separated the spectra of all components of the system, composed
of a binary system, whose southern component is partly resolved in the
H and K bands, plus an extended envelope, clearly detected both in
MIDI and L-band NACO data. We found that, while the extended
envelope contains most of the silicate (warm dust) emission at
10$\mu$m, the southern component can be well-described by a black-body
emission at 1300\,K. The temperature (1300\,K) and size (6\,AU at
1.5\,kpc) of this southern component is compatible with the inner-rim
emission of a circumstellar disk around a hot (B-type) star. On the
other hand, the northern component keeps its secrets by exhibiting an
unresolved shape with a variety of dust temperatures (from 1300\,K to
300\,K). It could be, for example, a T-Tauri star still deeply
embedded in its dust cocoon.

\section{Wide-field imaging: binarity at work?}

We also retrieved and reduced unpublished data of HD87643 from the WFI
camera in 2001. The image is shown in the bottom of
Fig.~\ref{figure_Hires_images}, together with a sketch of all the features
detected. In comparison with previous works
\citep{1972PASP...84..594V, 1981A&A....93..285S, 1983A&A...117..359S},
the image presented here has a larger dynamic range.

It appears that the nebula around HD87643 is made of three components:
filamentary structures, composing the main nebula in the north-west
quadrant (thick black line in the sketch); apparently blown-up
structures appearing as `` knots'' (labelled (B1) to (B4)), connected
to the previous filaments; and finally, arc-like structures (labelled
(A1) to (A7)), grouped in two sets south-east and 
south-west of the star, respectively.

The nebular filamentary structures can be explained by a past outburst
that took place $\approx$355\,yrs ago, given an expansion velocity of
$\approx1000$\,km\,s$^{-1}$ and a distance of 1.5\,kpc. The knots seen
in our image would correspond to denser interstellar clouds or clumps
that would offer more resistance to the nebular ejecta.

The arc-like structures appear regularly spaced in our image. At the
same adopted distance, they would correspond to regular ejections
every $\approx$14\,yrs to $\approx$50\,yrs, depending on the arc.
These broken structures suggest short, localised ejection that
might coincide with short periastron passages of the previously
detected companion, triggering violent mass-transfer between the
components.

\section{Comparison with previous works}

Previous observations of HD87643 did not detect the companion
star. Spectro-polarimetric \citep{1998MNRAS.300..170O} and
spectro-astrometric observations \citep{2006MNRAS.367..737B} had
the highest spatial resolution at that time. Both techniques provided
evidence for a significant north-south asymmetry of the system, but
the complexity of the signal prevented a direct interpretation, and
especially the detection of the binary star.

\begin{figure}[htbp]
  \centering
  \begin{tabular}{cc}
    \includegraphics[height=0.36\textheight]{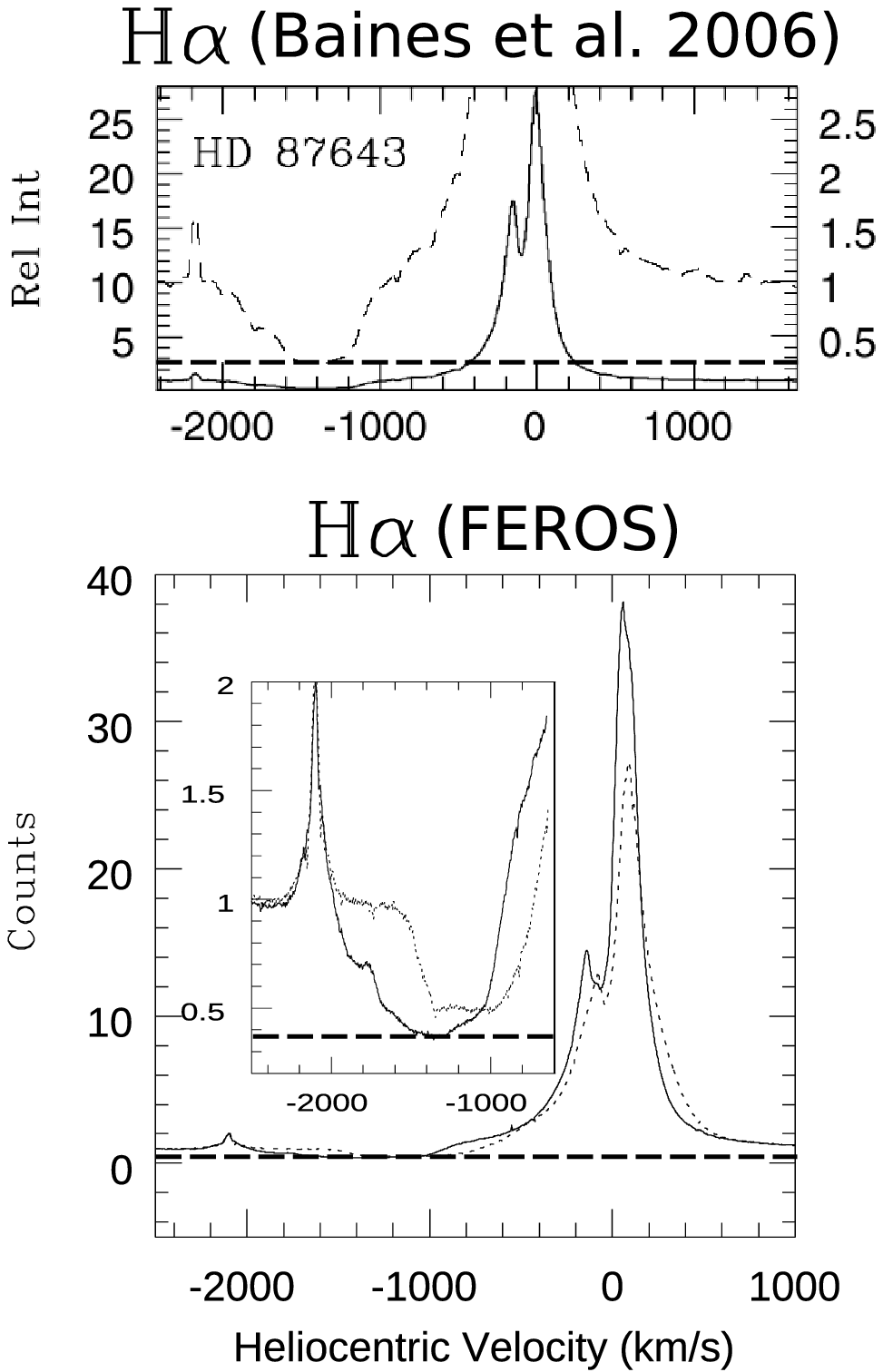}&
    \includegraphics[height=0.36\textheight]{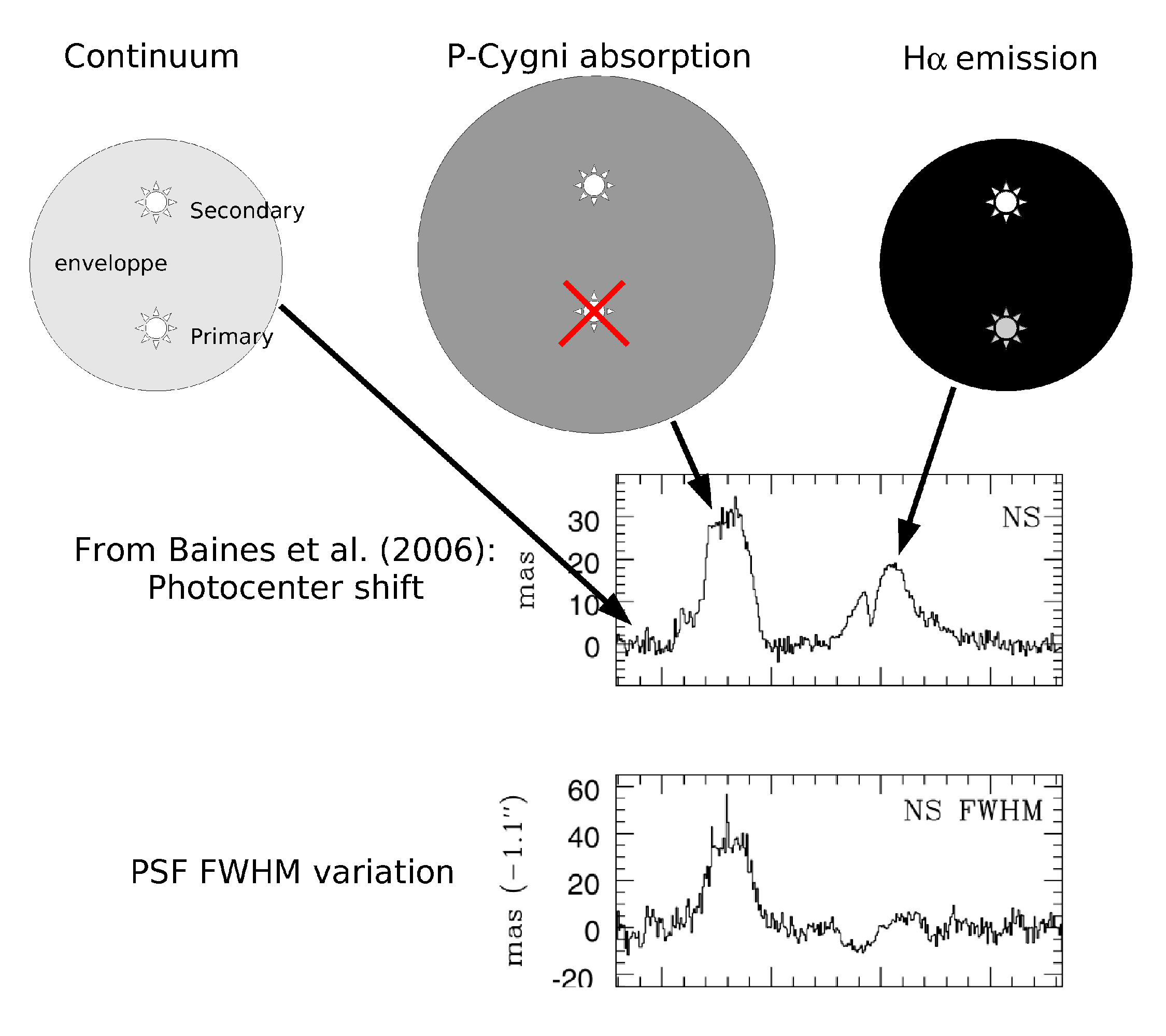}
  \end{tabular}
  \caption{{\bf Left:} H$\alpha$ line as seen by
    \citet[][top]{2006MNRAS.367..737B} and FEROS (1999, solid line and
    2000, dotted line). Dashed lines indicate the flat-bottomed shape
    of the P Cygni absorption.
    {\bf Right:} A tentative new interpretation of the
    \citet{2006MNRAS.367..737B} spectro-astrometric measurements: Both
    stars emits H$\alpha$, while only one is absorbed in the P Cygni
    absorption. Note the change in flux-balance of the two components
    between the continuum and both H$\alpha$ emission and P Cygni
    absorption.
    \label{fig:hydro}
  }
\end{figure}

The H$\alpha$ P-Cygni profiles of HD87643 (between -1000 and
-2000km\,s$^ {-1}$) show a flat-bottomed shape, seen in both of the
\citet{2006MNRAS.367..737B} and \citet{1998MNRAS.300..170O} spectra,
as well as in spectra we acquired with the FEROS instrument in 1999 and
2000 (see Fig.~ \ref{fig:hydro}). It could mean that in that spectral
region, {\it one} source of continuum is  strongly absorbed, while
other emission sources are not. This (variable) absorption represents
50\% to 60\% of the continuum.

The spectro-astrometric results from \citet{2006MNRAS.367..737B} can
be summarized as follows. The
photo-centre of the H$\alpha$ emission and of the P Cygni
absorption is shifted toward the north compared to the continuum
photo-center. The PSF FWHM is increased by 40mas both in the EW and NS
directions in the P Cygni absorption, but not in other
parts of the H$\alpha$ line. This FWHM increase is larger than our
inferred binary separation (34.5\,mas). This implies that the
remaining continuum originates from an extended component, unaffected
by the P Cygni absorption. This component contributes {\it at most}
(i.e. considering a fully saturated P Cygni absorption) 40-50\% of the
total visible continuum. We can tentatively attribute most of this
contribution to the scattered light from the circumbinary envelope,
detected by MIDI and NACO in the mid-infrared.

The photo-centre shifts may be understood as follows. The P-Cygni
absorption hides the southern component; hence, a northern
photo-centre shift is observed. Both stars emit H$\alpha$, but the
northern star emits more; so again, a northern photo-centre shift in
the H$\alpha$ emission is observed. In the P-Cygni absorption, the
large envelope surrounding the system appears larger as a consequence
of, e.g., increase of optical depth, leading to the PSF FWHM
increase. Therefore, in this frame, the spectro-astrometric
observations of \citet{2006MNRAS.367..737B} strongly support a
multiple-component origin for the H$\alpha$ emission.  This also
suggests that the northern companion is also a hot source able to
ionize circumstellar hydrogen.

\section{Discussion and conclusion}

We presented high angular resolution AMBER and NACO images as
well as a high dynamic range WFI image of HD87643. AMBER and NACO
images undoubtedly reveal a binary companion to the main star, while
the WFI one suggest a high-eccentricity orbit for the binary. Separating
the spectra of the different components in the system using the AMBER
and MIDI data, it was possible to infer that the main source is likely
a hot star encircled by a 6\,AU hot dust envelope, that the secondary is
embedded in a compact cocoon of dust, and that a circumbinary
envelope holds most of the dust silicate emission in the
system. Finally, we propose a new interpretation of literature
spectro-astrometric data, in which both stellar components emit
H$\alpha$.

The global view of the system has been completely changed by our
discovery of a companion to HD87643. The system might resemble in fact
the following: a hot star encircled by a dusty disk, whose inner rim
is seen in the AMBER image (a Herbig star?), a T Tauri companion
star, and a dusty circumbinary envelope. All this suggests a much
younger evolutionary status of HD87643 and, hence, a much closer
distance than previously thought. A monitoring campaing of the binary
throughtout its orbit would enable us to set accurately and
definitively the distance of the system and, hence, its evolutionary
status.





\begin{thebibliography}{9}
\expandafter\ifx\csname natexlab\endcsname\relax\def\natexlab#1{#1}\fi

\bibitem[{{Baines} {et~al.}(2006){Baines}, {Oudmaijer}, {Porter}, \&
  {Pozzo}}]{2006MNRAS.367..737B}
{Baines}, D., {Oudmaijer}, R.~D., {Porter}, J.~M., \& {Pozzo}, M. 2006, \mnras,
  367, 737

\bibitem[{{Leinert} {et~al.}(2004){Leinert}, {van Boekel}, {Waters},
  {Chesneau}, {Malbet}, {K{\"o}hler}, {Jaffe}, {Ratzka}, {Dutrey}, {Preibisch},
  {Graser}, {Bakker}, {Chagnon}, {Cotton}, {Dominik}, {Dullemond},
  {Glazenborg-Kluttig}, {Glindemann}, {Henning}, {Hofmann}, {de Jong},
  {Lenzen}, {Ligori}, {Lopez}, {Meisner}, {Morel}, {Paresce}, {Pel},
  {Percheron}, {Perrin}, {Przygodda}, {Richichi}, {Sch{\"o}ller}, {Schuller},
  {Stecklum}, {van den Ancker}, {von der L{\"u}he}, \&
  {Weigelt}}]{2004A&A...423..537L}
{Leinert}, C., {van Boekel}, R., {Waters}, L.~B.~F.~M., {et~al.} 2004, \aap,
  423, 537

\bibitem[{{Millour} {et~al.}(2009){Millour}, {Chesneau}, {Borges Fernandes},
  Meilland, Mars, Benoist, Thi\'ebaut, Stee, Hofmann, Baron, Young, Bendjoya,
  Carciofi, de~Souza, Driebe, Jankov, Kervella, Petrov, Robbe-Dubois, Vakili,
  Waters, \& Weigelt}]{Millour09}
{Millour}, F., {Chesneau}, O., {Borges Fernandes}, M., {et~al.} 2009, A\&A,
  accepted.

\bibitem[{{Oudmaijer} {et~al.}(1998){Oudmaijer}, {Proga}, {Drew}, \& {de
  Winter}}]{1998MNRAS.300..170O}
{Oudmaijer}, R.~D., {Proga}, D., {Drew}, J.~E., \& {de Winter}, D. 1998,
  \mnras, 300, 170

\bibitem[{{Petrov} {et~al.}(2007){Petrov}, {Malbet}, {Weigelt}, {Antonelli},
  {Beckmann}, {Bresson}, {Chelli}, {Dugu{\'e}}, {Duvert}, {Gennari},
  {Gl{\"u}ck}, {Kern}, {Lagarde}, {Le Coarer}, {Lisi}, {Millour}, {Perraut},
  {Puget}, {Rantakyr{\"o}}, {Robbe-Dubois}, {Roussel}, {Salinari}, {Tatulli},
  {Zins}, {Accardo}, {Acke}, {Agabi}, {Altariba}, {Arezki}, {Aristidi},
  {Baffa}, {Behrend}, {Bl{\"o}cker}, {Bonhomme}, {Busoni}, {Cassaing},
  {Clausse}, {Colin}, {Connot}, {Delboulb{\'e}}, {Domiciano de Souza},
  {Driebe}, {Feautrier}, {Ferruzzi}, {Forveille}, {Fossat}, {Foy},
  {Fraix-Burnet}, {Gallardo}, {Giani}, {Gil}, {Glentzlin}, {Heiden},
  {Heininger}, {Hernandez Utrera}, {Hofmann}, {Kamm}, {Kiekebusch}, {Kraus},
  {Le Contel}, {Le Contel}, {Lesourd}, {Lopez}, {Lopez}, {Magnard}, {Marconi},
  {Mars}, {Martinot-Lagarde}, {Mathias}, {M{\`e}ge}, {Monin}, {Mouillet},
  {Mourard}, {Nussbaum}, {Ohnaka}, {Pacheco}, {Perrier}, {Rabbia}, {Rebattu},
  {Reynaud}, {Richichi}, {Robini}, {Sacchettini}, {Schertl}, {Sch{\"o}ller},
  {Solscheid}, {Spang}, {Stee}, {Stefanini}, {Tallon}, {Tallon-Bosc}, {Tasso},
  {Testi}, {Vakili}, {von der L{\"u}he}, {Valtier}, {Vannier}, \&
  {Ventura}}]{2007A&A...464....1P}
{Petrov}, R.~G., {Malbet}, F., {Weigelt}, G., {et~al.} 2007, \aap, 464, 1

\bibitem[{{Rousset} {et~al.}(2003){Rousset}, {Lacombe}, {Puget}, {Hubin},
  {Gendron}, {Fusco}, {Arsenault}, {Charton}, {Feautrier}, {Gigan}, {Kern},
  {Lagrange}, {Madec}, {Mouillet}, {Rabaud}, {Rabou}, {Stadler}, \&
  {Zins}}]{2003SPIE.4839..140R}
{Rousset}, G., {Lacombe}, F., {Puget}, P., {et~al.} 2003, in SPIE Conf., Vol.
  4839, 140

\bibitem[{{Surdej} {et~al.}(1981){Surdej}, {Surdej}, {Swings}, \&
  {Wamsteker}}]{1981A&A....93..285S}
{Surdej}, A., {Surdej}, J., {Swings}, J.~P., \& {Wamsteker}, W. 1981, \aap, 93,
  285

\bibitem[{{Surdej} \& {Swings}(1983)}]{1983A&A...117..359S}
{Surdej}, J. \& {Swings}, J.~P. 1983, \aap, 117, 359

\bibitem[{{van den Bergh}(1972)}]{1972PASP...84..594V}
{van den Bergh}, S. 1972, \pasp, 84, 594

\end{thebibliography}

\end{document}